\documentclass[aps,prl,twocolumn]{revtex4}
\usepackage{amsmath}
\usepackage{graphicx,epsfig,psfrag}
\usepackage{amssymb}
\newcommand{\OLD}[1]{{\bf OLD RM}}

\newcommand{\IM}{{\text{Im}}}
\renewcommand{\vec}[1]{{\mathbf #1}}

\begin{document}
%


\newcommand{\anfz}[1]{\lq #1\rq{}}
\newcommand{\aanfz}[1]{\lq\lq #1\rq\rq{}}
\newcommand{\annih}[2][a]{\hat{#1}_{#2}}
\newcommand{\creat}[2][a]{\hat{#1}^\dagger_{#2}}
\newcommand{\textdef}[1]{\emph{#1}}
\newcommand{\intshort}[2][]{\int_{\mathbf{#2}_{#1}}}
\newcommand{\iintshort}[1]{\iint_{\mathbf{#1}_1,\mathbf{#1}_2}}
\newcommand{\sumshort}[2][]{\sum_{\mathbf{#2}_{#1}}}
\newcommand{\ssumshort}[1]{\sum_{\mathbf{#1}_1,\mathbf{#1}_2}}
\newcommand{\hannih}[3][a]{\hat{#1}^{\text{(H)}}_{#2}(#3)}
\newcommand{\hcreat}[3][a]{\hat{#1}^{\dagger\text{(H)}}_{#2}(#3)}
\newcommand{\inproduct}[4]{\mathbf{#1}_{#2}\cdot\mathbf{#3}_{#4}}
\newcommand{\op}[1]{\hat{{\cal #1}}}
\newcommand{\tr}{\operatorname{tr}}
\newcommand{\ad}{\operatorname{ad}}
\newcommand{\Ad}{\operatorname{Ad}}
\newcommand{\sgn}{\operatorname{sgn}}
\newcommand{\sdet}{\operatorname{sdet}}
\newcommand{\str}{\operatorname{str}}
\newcommand{\bra}[1]{\langle #1 \vert}
\newcommand{\ket}[1]{\vert #1 \rangle}
\newcommand{\braket}[2]{\bra{#1} #2 \rangle}
\newcommand{\ham}{\hat{H}}
\newcommand{\dens}{\hat{\rho}}
\newcommand{\meas}[1]{d\left[ \bar{#1}, #1 \right]}
\newcommand{\rdbra}[1]{\left( #1 \right)}
\newcommand{\slbra}[1]{\left\{ #1 \right\}}
\newcommand{\hkbra}[1]{\left[ #1 \right]}
\newcommand{\rhoeff}{\dens_{\text{eff}}}
\newcommand{\order}[1]{{\cal O}\rdbra{#1}}
\newcommand{\orderesq}{{\cal O}(\epsilon^2)}
\newcommand{\fieldpair}[2][]{\bar{#2}_{#1},#2_{#1}}
\newcommand{\fielddep}[2][]{(\fieldpair[#1]{#2})}
\newcommand{\functdep}[2][]{[\fieldpair[#1]{#2}]}
\newcommand{\functint}[2][]{\int{\cal D}\functdep[#1]{#2}\,}
\newcommand{\sumprime}[1]{\sideset{}{'}\sum_{#1}}
\renewcommand{\Re}{\operatorname{Re}}
\renewcommand{\Im}{\operatorname{Im}}
\newcommand{\diag}{\operatorname{diag}}
\newcommand{\opone}{\leavevmode\hbox{\small1\kern-3.3pt\normalsize1}}
\newcommand{\average}[2][]{\bigl< #2 \bigr> _{#1}}
\newcommand{\sub}[1]{_{\text{#1}}}
\newcommand{\super}[1]{^{\text{#1}}}
\newcommand{\abs}[1]{\left\vert #1 \right\vert}
\newcommand{\symbr}[1]{\sigma_3^{\text{#1}}}
\newcommand{\vol}[2][]{\operatorname{vol}_{#1}#2}
\newcommand{\opn}[1]{\operatorname{#1}}
\newcommand{\mb}[1]{\mathbf{#1}}
\newcommand{\diff}{\operatorname{d}\negthinspace}
\newcommand{\swok}[1][n]{\hat{\Delta}^j(\mathbf{#1})}
\newcommand{\red}[1]{\textcolor{red}{#1}}
\newcommand{\redbox}[2][2pt]{
\begin{center}
\setlength{\fboxrule}{#1}\fcolorbox{red}{white}{\mbox{#2}}
\end{center}}
\newcommand{\entspricht}{$\overset{\wedge}{=}$}
\newcommand{\ruskij}{\i\u{\i}}


%
\title{Universal dephasing rate due to diluted Kondo impurities}
\author{T. Micklitz$^1$, A. Altland$^1$, T. A. Costi$^2$, and A. Rosch$^1$ }
\affiliation{$^1$ Institute for Theoretical Physics, University of Cologne, 50937
Cologne, Germany \\
$^2$ Institut f\"ur Festk\"orperforschung, Forschungszentrum J\"ulich,
52425 J\"ulich, Germany.}

\date{\today}
\newcommand{\tp}{\tau_\varphi}
\begin{abstract}
  We calculate the dephasing rate due to magnetic impurities in a
  weakly disordered metal as measured in a weak localization
  experiment.  If the density $n_{\text{S}}$ of magnetic impurities is
  sufficiently low, 
  the dephasing rate $1/\tau_\varphi$ is a universal function, $1/\tau_\varphi
  = ( n_{\text{S}}/\nu) f\!\left(T/T_{\text{K}}\right)$, where $T_{\text{K}}$ is the
  Kondo  temperature and $\nu$ the density of states. We show that
  inelastic vertex corrections with a typical energy transfer $\Delta
  E$ are suppressed by powers of $1/(\tau_\varphi \Delta E) \propto n_{\text{S}}$.
  Therefore the dephasing rate can be calculated from the {\em
    inelastic cross section} proportional to $\pi \nu \IM T- |\pi \nu
  T|^2$, where $T$ is the $T$-matrix which is evaluated numerically
  exactly using the numerical renormalization group.
\end{abstract}
\maketitle
Dephasing, i.e. the loss of wave coherence, is a ubiquitous phenomenon in
the quantum mechanics of complex systems. It is of relevance to any
experiment where both interference and interactions play a role and
is, therefore, of profound importance in all areas of nanoscopic and
mesoscopic physics.

While the basic phenomenon of dephasing as such is of rather general
nature, the concrete definition of a dephasing rate, and its
experimental determination vary from context to context. In
this Letter we consider the dephasing rate as determined by
weak-localization (WL) measurements in metals \cite{AlAr85}. In weakly
disordered metals, the interference of electronic wave functions on
time-reversed paths leads to a characteristic reduction (or
enhancement in the presence of spin-orbit interactions) of the
conductivity. The magnitude of this effect is
controlled by the dephasing time $\tp$ --- the typical
time-scale over which electrons get entangled with their environment
(phonons, other electrons or dynamical impurities) thereby loosing the
ability to interfere.  Even small magnetic fields break time-reversal
invariance thus prohibiting the interference of time-reversed
paths. Fitting the magnetoresistivity to WL theory
is a means to determine the dephasing rate $1/\tp$ with high precision.

Surprisingly, most of these experiments \cite{MoJaWe97}
show a saturation of the
dephasing rate at the lowest 
accessible temperatures
$T$, while theoretically it is expected that in the limit $T\to 0$ all
inelastic processes freeze out when the system approaches its
(time-reversal invariant and unique) ground-state. This has lead to an
intense discussion \cite{GoZa98, AlAlGe99,
  De03,MoJaWe97} as to whether quantum fluctuations can induce
dephasing at $T=0$. While we believe that this latter scenario is theoretically
excluded for electrons in a disordered metal, the presence of only a few parts-per-million (ppm) of dynamical
impurities --- realized by atomic two-level systems \cite{ZaDeRa99} or
by magnetic impurities \cite{MoWe00, HaVrBr87,ScBaRaSa03,PiGoAnPoEsBi03} --- may be
an alternative cause of the saturation phenomenon. Indeed it has
been shown experimentally that  magnetic impurities lead
to an apparent saturation of $1/\tp$ at least in some $T$ range
\cite{ScBaRaSa03}. In contrast, some extremely pure Au and Ag samples with
negligible concentration of impurities \cite{PiGoAnPoEsBi03} show no
saturation and seem to follow the predictions of Altshuler, Aronov and
Khmelnitsky \cite{AlArKh82} for the dephasing induced by Coulomb
interactions.

The effect of dynamic magnetic impurities on $1/\tp$ has first been
considered by Ohkawa, Fukuyama and Yosida \cite{OhFuYo83} (for the static case see Ref.~\cite{HiLaNa80})
using perturbation theory (generalized to renorma\-lized perturbation theory in Refs.~\cite{HaVrBr87,VaGlLa03})
which limits the range of
applicability to temperatures $T \gg T_K$ larger than the Kondo
temperature, $T_K$. For $T\ll T_K$, a quadratic $T$ dependence,
$1/\tp \propto T^2$ has been predicted \cite{VaGlLa03} based
on Fermi liquid arguments. 
In this paper we argue that the impact of static 
and diluted dynamic impurities can largely be
treated separately. This separation enables us to combine a
perturbative approach to the static disorder with a (numerically)
exact treatment of the Kondo interaction. In this way, the effect of
a small concentration, $n\sub{S}$, of magnetic impurities on $1/\tp$
can be explored in the entire crossover range
 from $T \gg T_K$ down to $T\ll T_K$.


We consider the Hamiltonian $H = H_0 + H\sub{S}$, where
\begin{align}\label{h_0}
H_0 = \int d^d\bold{x}\, c^\dagger_\sigma(\bold{x}) \Big[ \frac{\hat{\bold{p}}^2}{2m} - \mu + V(\bold{x}) \Big] c_\sigma(\bold{x})
\end{align} describes electrons in a weak non-magnetic disorder
potential $V$, modeled for convenience by Gaussian white noise:
$\langle V(\bold{x})V(\bold{x'})\rangle_V = \frac{1} {2\pi\nu\tau}
\delta(\bold{x}-\bold{x'})$ where $\nu$ is the density of states
(DOS), $\tau=l/v\sub{F}$ the elastic scattering time and $l$ the
elastic mean free path. (None of our results is affected by the
precise choice of band-structure or model of disorder.) The coupling
of the electrons to a small concentration $n\sub{S}$ of spin-$1/2$
impurities is described by the Kondo Hamiltonian
\begin{align}\label{h_s}
H\sub{S} = J \sum_i \hat{\bold{S}}_i c^\dagger_\sigma(\bold{x}_i)  \bold{\sigma}_{\sigma\sigma'} c_{\sigma'}(\bold{x}_i).
\end{align}
To calculate physical quantities, we have (i) to average
over $V(\bold{x})$ and (ii) the positions $\bold{x}_i$ of
the spins, taking into account (iii) the exchange coupling $J$ to all
orders. Two small parameters will help us in attacking this
problem: The concentration of magnetic impurities $n\sub{S}$ is tiny
[more precisely $n\sub{S}/(\nu T\sub{K}) \ll 1$] and also the ratio of
electronic wave-length and elastic mean free path, $1/(k\sub{F} l)$,
is small but finite.

\begin{figure}[t]
\centering \includegraphics[height=3.5cm,width=7.5cm]{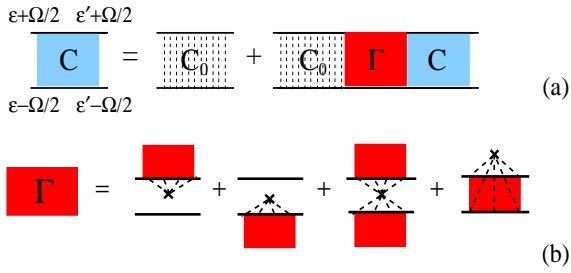}
\caption{\label{fig1} Bethe-Salpeter equation for the Cooperon $C$ in
  the presence of (dilute) magnetic impurities to linear order in $n\sub{S}$. $C_0$ is the bare
  Cooperon in the absence of interactions and $\Gamma$ the irreducible
  vertex obtained by adding self-energy, elastic- and inelastic vertex
  contributions. The crosses with attached dashed lines denote the averaging
over impurity positions $\vec{x}_i$, the squares the interactions to arbitrary order in $J$.}
\end{figure}

We wish to explore the impact of dynamic impurity scattering on the
WL corrections to the electric conductivity.  Technically, this
amounts to computing the impurity generated 'self energy' or 'mass'
of the Cooperon describing the coherent propagation of an electron
and a time-reversed hole in the disordered environment. As justified
in detail below, we neglect mixed interaction/disorder diagrams. The
problem therefore reduces to the solution of the Bethe-Salpeter
equation depicted diagrammatically in Fig.~\ref{fig1}(a).  To linear
order in $n\sub{S}$, the two-particle irreducible vertex $\Gamma$
shown in Fig.~\ref{fig1}(b) can be separated into three distinct
contributions: self-energy diagrams (the first two terms in
Fig.~\ref{fig1}(b)), an `elastic' vertex correction with no energy
transfer between upper and lower line, and an `inelastic' vertex
where interaction lines connect the two lines. We begin by focusing
on the elastic contributions to the Cooperon self energy as
inelastic contributions give vanishingly
small corrections for small
$n\sub{S}$, see below.

Since self-energy and elastic vertex contributions conserve the energy
of single electron lines, the solution of the reduced Bethe-Salpeter
amounts to a straightforward summation of a geometric series.  Setting
the center-of-mass frequency $\Omega$ (see Fig.~\ref{fig1}) to $0$,
the Cooperon obtains as $ C^{\bold{q}}_{\Omega=0} (\epsilon, \epsilon')
= \frac{1} {2\pi\nu\tau^2} \delta (\epsilon- \epsilon') /[D
\bold{q}^2+1/ \tp(\epsilon,T)]$ and the WL correction
to the conductivity is given by
\begin{align}\label{wl}
\frac{\Delta \sigma_{\rm WL}}{\sigma_{\rm Drude}}= 2\tau^2\int d
\epsilon f'(\epsilon) \int d^d \vec{q}  \frac{1}{D
  \bold{q}^2+1/\tp(\epsilon)},
\end{align}
with the $T$ and $\epsilon$ dependent dephasing rate
\begin{align}\label{tau}
\frac{1}{\tp(\epsilon,T)} = \frac{2n\sub{S}}{\pi\nu} \left[ \pi\nu
  \opn{Im}\!\!\left[ \opn{T}\super{A}(\epsilon) \right] - |\pi\nu
  \opn{T}\super{R}(\epsilon)|^2 \right].
\end{align}
The many-body $\opn{T}$-matrix [defined by the Green function
$G_{\bold{x}\bold{x'}}(\epsilon)=G^0_{\bold{x}\bold{x'}}(\epsilon)+G^0_{\bold{x}\bold{0}}(\epsilon)T(\epsilon)G^0_{\bold{0}\bold{x'}}(\epsilon)$]
describes the scattering of electrons from a {\em single} magnetic
impurity. Eq.~(\ref{tau}) affords a simple interpretation:
$\opn{Im}\!\!\left[ \opn{T}\super{A}(\epsilon) \right] $ is
proportional to the total cross section, while
$|\opn{T}\super{R}(\epsilon)|^2$ describes the elastic cross
section. Their difference is -- by definition -- proportional to the
{\em inelastic} cross section recently been introduced by Zarand {\it
  et al.} \cite{ZaBoDeAn04}. Note that the optical theorem guarantees
the vanishing of the inelastic cross-section for static impurities.
Together with the numerical evaluation of $1/\tp$ and the 
analytic estimates of the leading corrections, given below, Eq.~(\ref{tau}) 
is the main result of this paper.

\begin{figure}[t]
\centering \includegraphics[width=7.5cm]{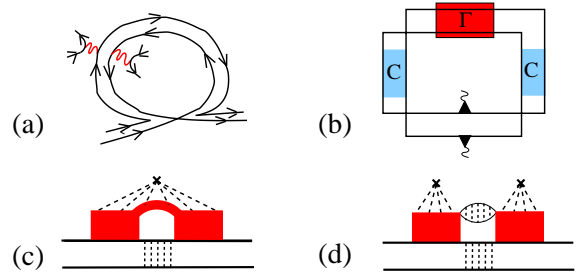}
\caption{\label{fig2}
a) Interference of interacting electrons. b) Corrections due to the inelastic
vertex to lowest order. c) WL correction to the dephasing rate
linear in $n\sub{S}$. d) Altshuler-Aronov-Khmelnitsky type corrections arising to
order $n\sub{S}^2$.}
\end{figure}

To determine the range of applicability of Eq.~(\ref{tau}), three
classes of corrections have to be considered: contributions from the
inelastic vertex, mixed interaction-disorder diagrams, and
higher-order corrections in $n\sub{S}$. We begin by considering the
lowest order expansion of the WL correction in the inelastic vertex
(cf. Fig.~\ref{fig2}(b)),
\begin{align}\label{wl1}
& \Delta\sigma\sub{1} = \int d\epsilon \int d\Omega \,\tilde{f}(\epsilon,\Omega) \Gamma\sub{in}(\epsilon,\Omega) I_d(\epsilon, \Omega).
\end{align}
Here, $\tilde{f}$ denotes some thermal function restricting
$\epsilon$ and $\Omega$ to values smaller than $T$, $\Gamma\sub{in}$
is the inelastic vertex and $I_d(\epsilon, \Omega) \sim
\left[(1/\tp)^2+\Omega^2 \right]^{\frac{d-4}{4}}$ results from
momentum-integration over two Cooperons in  $d$ dimensions. For
sufficiently small  $n\sub{S}\ll \nu T_K$ the typical energy $\Delta
E \sim \Omega\sim T$ ~\cite{fnTtrans} exchanged at the inelastic
vertex greatly exceeds the dephasing rate (see below) $1/\tp\propto
n\sub{S}$. Evaluating the integral~\eqref{wl1} for $\Delta E \gg
1/\tau_\phi$, we obtain 
\begin{align}\label{s1}
\frac{\Delta\sigma\sub{1} }{\Delta\sigma_{\rm WL}} \sim \left( \frac{1}{\Delta E
\,  \tp }\right)^{(4-d)/2} \lesssim    \left( \frac{n\sub{S}}{\nu
T\sub{K} }\right)^{(4-d)/2}  \ll 1.
\end{align}
It is straightforward to verify that
this estimate pertains to higher orders in the expansion in the
inelastic vertex.

The irrelevancy of the inelastic vertex for processes with typical
energy transfer larger than $1/\tp$ has been observed before by
several authors (e.g. \cite{AlArKh82,MeFaAl02,VaGlLa03}): Semiclassically, 
vertex corrections describe the interference of two
(time-reversed) electrons undergoing an interaction process (see
Fig.~\ref{fig2}(a)). Due to the transferral of a certain energy
$\Delta E$, the two electrons subsequently accumulate a phase
difference, $e^{i \Delta E \, t}$, where $t\sim \tp$.
Therefore, vertex corrections with $\Delta E>1/\tp$ (characteristic
for our problem) do not effectively contribute to WL. (In
contrast, Coulomb interactions in $d\le 2$ are dominated by
low-energy transfers and vertex corrections are 
required to regularize infrared singularities~\cite{AlArKh82}.)

We next turn to the discussion of the second family of corrections
potentially altering our above results, correlated
disorder-interaction processes as shown in Fig.~\ref{fig2}(c).  A
preliminary indication as to the relevance of such contributions may
be obtained by estimating the sample-to-sample fluctuations~\cite{kettemann} of the Kondo
temperature, $T\sub{K}$,
\begin{align}\label{tk1} \Big(
  \frac{\delta T\sub{K}}{T\sub{K}}\Big)^2 = \frac{1}{\nu^2}
  \int_{T\sub{K}}^{E\sub{F}} d\omega \int_{T\sub{K}}^{E\sub{F}} d\omega'
  \frac{\big\langle\delta\nu(\omega)\delta\nu(\omega')\big\rangle_V}{\omega\omega'}.
\end{align}
Substituting the results for the fluctuations of the local density of
states $\delta\nu$ in weakly disordered $d$--dimensional metals
\cite{AlAr85}, we obtain
\begin{align}\label{tk2}
\left( \frac{\delta T\sub{K}}{T\sub{K}}\right
)^2 \sim
\begin{cases}
\frac{1}{(k\sub{F}L_\bot)^2} \frac{1}{\sqrt{\tau T\sub{K}}} & \quad \text{in (quasi) $d=1$,} \\
\frac{1}{k\sub{F}l} \frac{1}{(J\nu)^3} & \quad \text{in $d=2$,} \\
\frac{1}{(k\sub{F}l)^2} \frac{1}{(J\nu)^2} & \quad \text{in $d=3$,}
\end{cases}
\end{align}
where $L_\perp^2$ is the cross-section of a quasi one-dimensional
wire. We assume, henceforth, that $k\sub{F} l$ is
sufficiently large, so that $\delta T\sub{K} \ll T\sub{K}$. While this
condition seems restrictive in quasi-$1d$, it turns out to
be always met in the WL regime, $\Delta\sigma\super{WL} \ll
\sigma\super{Drude}$, realized in experiments (and assumed in this
paper).

More formally, the role of correlations disorder/interactions may be
explored in terms of the diagrams shown in Fig.~\ref{fig2}(c). On the
face of it, these diagrams are smaller by factors of $1/(k_F l)$ than
the leading contributions considered above (as quantum interference maintained across the impurity limits the momentum exchanged  to values
$\lesssim l^{-1}$ much smaller than $ k_F$).
However, for very low $T$ the enhanced
infrared singularity caused by the presence of extra diffusion modes
may over-compensate this phase space suppression factor. Using that
for $T\ll T_K$, the bare interaction may be described by Fermi liquid
theory \cite{hewson}, we find that only in one dimension the
additional diagrams ($\propto T^{(d+2)/2}+T^2$) lead to contributions of anomalously strong
singularity. Specifically, we obtain a correction to the dephasing
rate,
$ 1/\tau_{\varphi,c}\sim n\sub{S} T^{3/2}/[\nu T\sub{K}^2 \sqrt{\tau} (k_F
  L_\perp)^2]$. Therefore, for
 \begin{align}
T \lesssim  \frac{1}{(k\sub{F}L_\bot)^4} \frac{1}{\tau
  T\sub{K}}T\sub{K} \ll T_K
\end{align}
the separation disorder/interactions used above becomes invalid in $d=1$.

At very low $T$, yet another type of correction begins to play a
role: The diluted Kondo impurities become indistinguishable from a
conventional disordered Fermi liquid with short-range
momentum-conserving interactions \cite{shortRangeInt} and the
dephasing rate is determined by Altshuler-Aronov-Khmelnitsky
\cite{AlArKh82,shortRangeInt} type processes which, in our context,
are encapsulated in the third family of diagrams shown in
Fig.~\ref{fig2}(d). Contributing only at order $n\sub{S}^2$ they
generate corrections scaling as $T^{2/3}$, $T$, and $T^{3/2}$ in
$d=1,2,3$, respectively.  Evaluating the prefactors (again in Fermi
liquid theory) we find that these contributions become sizeable at
temperatures
\begin{align}
T \lesssim \left\{ \begin{array}{ll}
\frac{1}{(k\sub{F}l)^4} \Big(\frac{n\sub{S}}{\nu T\sub{K}}\Big)^2\tau T\sub{K}^2  & d=3,\\[1mm]
 \frac{1}{k\sub{F}l} \frac{n\sub{S}}{\nu T\sub{K}} T\sub{K}& d=2, \\
 \frac{1}{k\sub{F}L_\bot} \frac{1}{(\tau T\sub{K})^{1/4}}\Big(\frac{n\sub{S}}{\nu T\sub{K}}\Big)^{1/4}T\sub{K}& d=1.
 \end{array}\right.
\end{align}
In all cases the crossover scale is well below $T\sub{K}$ and may
arguably be neglected in all relevant experiments. Further corrections
to $1/\tp$ of order $n\sub{S}^2$ and higher arise from clusters of two and
more magnetic impurities which are sufficiently close that the
inter-impurity coupling dominates over the Kondo effect
\cite{spinGlasPaper}.

Fitting to $\Delta
\sigma = 2\tau^2 \sigma_\mathrm{Drude} \int d^d\mathbf{q}
(D\mathbf{q}^2+\tp^{-1}(T))^{-1}$,
we next relate our results to the $T$--dependent dephasing
rates, $\tp^{-1}(T)$ extracted from experiments. Comparison with
the energy--resolved representation \eqref{wl} results in
\begin{align}
\frac{1}{\tp(T)} =\left\{
\begin{array}{ll}
 \left[ - \int d\epsilon f'(\epsilon)
\tp(\epsilon,T)^{\frac{2-d}{2}} \right]^{\frac{2}{d-2}} & d=1,3, \\[2mm]
\exp\!\left[ \int d\epsilon f'(\epsilon)
\ln \tp(\epsilon,T)\right] & d=2,\\[2mm]
 - \int d\epsilon f'(\epsilon)
/\tp(\epsilon,T) & \omega_B \tp \gg 1, \\
\end{array}\right.\nonumber
\end{align}
where the last line applies  to the case where a strong magnetic field
is present and the Cooperon can be expanded in
$1/(\omega_B \tp)$. ($\omega_B$ is the ''cyclotron'' frequency of the
Cooperon.)

\begin{figure}[t]
\centering \includegraphics[width=0.88
\linewidth,clip=]{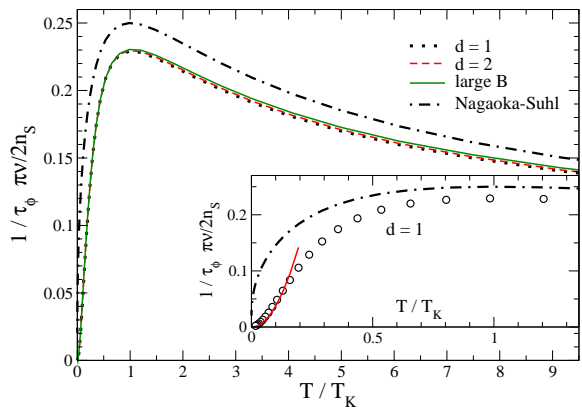}
\caption{\label{fig3} Universal dephasing rate calculated via NRG.
Accidentally, there is almost no dependence on the dimension $d$.
The dot-dashed line shows the result of a Nagaoka-Suhl resummation \cite{correct}, $\frac{1}{\tp} =\frac{n\sub{S}}{2 \pi\nu} \frac{\pi^2 3/4}{\pi^2 3/4+\ln^2 T/T\sub{K}} $, using our definition of $T\sub{K}$ \cite{tk}.  While the Fermi liquid behavior is recovered for $T \lesssim 0.1
T\sub{K}$, the calculated prefactor of the $T^2$ behavior (solid line)
is not exactly reproduced due to numerical problems \cite{costi}. }
\end{figure}

Just two parameters enter the expression for $1/\tp$: the Kondo
temperature $T\sub{K}$ and the dimensionless density of magnetic
impurities $n\sub{S}/(\nu T\sub{K})\ll 1$.  The $T$-matrix in
Eq.~\ref{tau} can be evaluated for arbitrary values of $T/T\sub{K}$
and $\epsilon/T\sub{K}$ using the numerical renormalization group
\cite{wilson} generalized to the calculation of dynamical quantities
in Refs.~\cite{costi,hofstetter}). The result is shown in Fig.~\ref{fig3} where we
define \cite{tk} and determine the Kondo temperature
$T\sub{K}=T\sub{K}^{(0)}$ from the $T=0$ susceptibility $\chi=(g
\mu_B)^2/(4 T\sub{K}^{(0)})$.  For $T\gg T_K$, we obtain  (correcting factors 2 in
  \cite{HaVrBr87} and 16 in \cite{VaGlLa03})  $1/\tp \approx  3 \pi n\sub{S}/(8 \nu
\ln^2[T/T_K])$ 
giving rise to a very weak $T$ dependence for $T\gtrsim T_K$. For 
$T\lesssim 0.3 \, T\sub{K}$ one observes a
crossover regime where $1/\tp$ varies almost linear in $T$ while at
lowest $T$ one obtains the expected $T^2$ behavior.  This
low-temperature regime can also be calculated analytically using Fermi
liquid theory \cite{hewson},
\begin{align}
\frac{1}{\tp(T)} \approx c_{\rm FL} \frac{2 n\sub{S}}{\pi\nu}  \left(\frac{T}{T\sub{K}}\right)^2 \frac{\pi^4}{24}
\end{align}
with $c_{\rm FL}\approx 0.927, 0.946, 0.969$ in $d=1,2,3$ and $c_{\rm
  FL}=1$ for $\omega_B \tp\gg 1$ (correcting prefactors in
   Ref.~\cite{VaGlLa03}).
Accidentally, the dependence of both the analytical and numerical
results on $\omega_B \tp$ (and on $d$) is very weak, implying that
the usual fits of the magnetoresistivity can be used to determine
$1/\tp$.


In the comparison to concrete experiments one needs to account for
the interplay of dephasing due to magnetic impurities and due to
Coulomb interactions \cite{AlArKh82}. Since the latter are
controlled by infrared divergences in $d\le 2$, the respective
rates do not add. Instead one needs to solve, e.g. in quasi
one-dimensional systems, the equation
\begin{align}\label{coulomb}
\frac{1}{\tp} =  & \, \kappa T \sqrt{\tp} + \frac{1}{\tau_{\varphi,\rm{S}}}
 \approx & \!
\begin{cases}
(\kappa T)^{2/3} + 2/(3 \tau_{\varphi,\rm{S}})\\
1/\tau_{\varphi,\rm{S}} + \kappa T \sqrt{\tau_{\varphi,\rm{S}}},
\end{cases}
\end{align} where the first term describes the  self-consistently calculated effects of Coulomb interactions while $1/\tau_{\varphi,\rm{S}}$ is
the dephasing rate due to the magnetic impurities. The first
(second) line holds when the Coulomb dephasing (Kondo dephasing)
dominates.

In conclusion, we have shown that the dephasing rate due to diluted
magnetic impurities can be calculated directly from the inelastic
cross section (\ref{tau}) introduced in \cite{ZaBoDeAn04}. This result
is valid for all types of diluted dynamical impurities as long as
typical energy transfers remain larger than $1/\tp$. In the case of
Kondo impurities the dephasing rate depends on just two parameters,
the Kondo temperature $T\sub{K}$ and the dimensionless density of
magnetic impurities $n\sub{S}/(\nu T\sub{K})\ll 1$.  
 A measurement of $\tp$ for
spin-$1/2$ impurities along with an independent experimental
determination of $T\sub{K}$ and $n\sub{S}$ would put our theory to a
parameter--free test [up to fitting the Coulomb background,
Eq.~(\ref{coulomb})]. An open question is how spin-orbit interactions and disorder influence  the Kondo effect in systems with larger spins, e.g. Fe in Au  \cite{ScBaRaSa03}.

We acknowledge 
discussions with
 Ch.~B\"auerle,  L.~Glazman, J.~v.~Delft, S.~Mirlin,
 B.~Spivak and P.~W\"olfle and
 financial support of the DFG by SFB 608 and TR~12.


\begin{thebibliography}{}
\bibitem{fnTtrans}
More precisely,
$\Delta E\approx T$ in the Fermi liquid regime $T\ll T\sub{K}$ while
for $T \gg T\sub{K}$, $\Delta E\approx T/\ln^2[T/T\sub{K}]$  is given by the
life-time of the spin. Comparing this with
$1/\tp\sim n\sub{S}/\nu$ at $T\sim T_K$, we obtain the
criterion stated in the text.
\bibitem{tk}  In literature \cite{hewson} a number of other definitions
of $T\sub{K}$ are used. The perturbative $T\sub{K}$  is given by $0.4107\,
T\sub{K}^{(0)}$  \cite{hewson}, the half width half maximum of the spectral function at $T=0$
is $\approx 2.05 \, T\sub{K}^{(0)}$. At $T \approx 0.94\, T\sub{K}^{(0)}$ ($T \approx 0.9\, T\sub{K}^{(0)}$)
the resistivity of diluted Kondo impurities in a dirty (clean) metal and the conductance through a quantum dot have half of their maximal value.
\end{thebibliography}
\end{document}